\documentstyle[aps,prl,multicol,epsf,graphicx]{revtex}
\begin{document}
\title{\bf
 Large Scale Structures, Symmetry, and Universality in Sandpiles}

\author{David Hughes and Maya Paczuski}
\address{Department of Mathematics,  Imperial College
of Science, Technology and Medicine, London UK SW7 2BZ}
\date{\today}
\maketitle

\begin{abstract}

We introduce a sandpile model where, at each unstable site, all
grains are transferred randomly to downstream neighbors.  The
model is local and conservative, but not Abelian.  This does not
appear to change the universality class for the avalanches in the
self-organized critical state.  It does, however, introduce
long-range spatial correlations within the metastable states. We
find large scale networks of occupied sites whose density vanishes
in the thermodynamic limit, for $d \geq 1$.

\end{abstract}


\begin{multicols}{2}

One of the puzzling questions about macroscopic complex phenomena
concerns the mechanisms responsible for the large spatially
correlated structures that are often seen in Nature.  It has been
proposed that Self-Organized Criticality (SOC)\cite{btw} may be
one mechanism, where the bursty, scale-free threshold dynamics of
a slowly driven system is intimately linked to the emergence of
long-range spatial (and temporal) correlations in it \cite{soc}.
An obvious candidate for this picture would be the stick-slip
dynamics of earthquakes, described by the Gutenberg Richter power
law distribution for seismic moments, and faults, which form a
fractal pattern in the crust of the earth.  However, the simple
sandpile, or earthquake models do not clearly show large scale
structures. Furthermore, although many macroscopic systems show
bursty transport phenomenology, a general feature of SOC, the link
is not yet established because questions of robustness and
universality are not yet resolved.  Since the number of possible
models that may be studied numerically is inexhaustible, it is
essential to determine the symmetry (or other) criteria for
universality \cite{universal} and robustness of SOC.  For this
purpose, we consider simple models that may be related to
analytically solvable ones.

\begin{figure}
\narrowtext
\includegraphics[width=230pt]{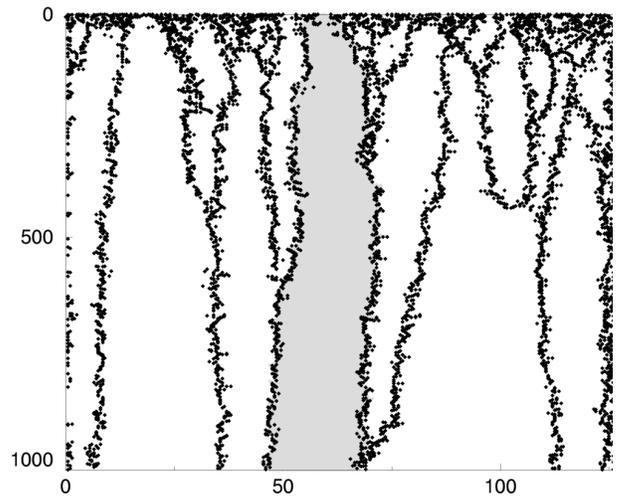}
{\vskip 10pt} \caption{A steady-state configuration of our $d=1$
sandpile model showing sites occupied with grains forming a
network that can transport grains from one end of the sandpile to
the other. The shaded area indicates the sites which toppled in
the preceding avalanche.} \label{fig:1+1dlattice}
\end{figure}

\begin{figure}
\narrowtext
\includegraphics[width=230pt]{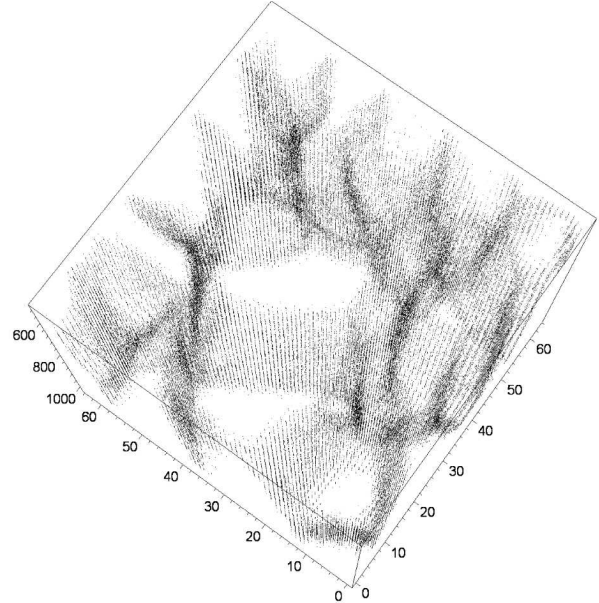}
{\vskip 10pt} \caption{A steady state configuration of the
sandpile model in $d=2$ dimensions. Note that the axes are not
scaled proportionally and lattice sites with $t<500$ have been
removed for clarity.} \label{fig:2+1dlattice}
\end{figure}

Here we propose what may be the simplest sandpile model that gives
large spatially correlated structures. For $d\geq 1$ the
avalanches in the model have a scale-free distribution with
critical coefficients in the same universality class as the
Abelian Stochastic Directed Sandpile Model (A-SDM)
\cite{pv,pb,maslov}.  There it has been proven that no spatial
correlations exist in the steady state metastable configurations
\cite{pb}.  The model we introduce is closely related to the
A-SDM. However, a change in the rule for updating unstable sites
breaks the Abelian symmetry.  (The Abelian symmetry refers to the
fact that the order for updating the unstable sites has no effect
on the final state that is reached.)  This symmetry breaking
introduces obvious large scale structures, consisting of networks
of occupied sites, within the metastable states that are reached
in the steady state, as shown in Figs.~\ref{fig:1+1dlattice} and
\ref{fig:2+1dlattice}.  These spatial correlations are not present
when the symmetry is restored.  The avalanches change the network
configuration slowly, just as earthquakes change the configuration
of faults slowly.  During a single or a few events it might appear
falsely that the configuration is static or ``pre-existing''.

Breaking the Abelian symmetry, however, has no effect on the
critical exponents for the avalanches, for $d \geq 1$.  There is
universality and robustness for the bursty phenomena with respect
to breaking the Abelian symmetry.  Thus two systems in the same
universality class with respect to the scaling behavior of
avalanches show totally different structures of the metastable
states, with one being completely uncorrelated and the other
having channels, or networks at large scales.

\begin{figure}
\narrowtext
\includegraphics[width=230pt]{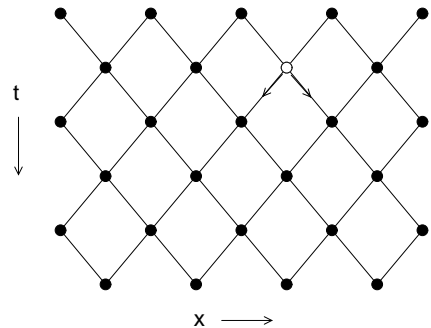}
\caption{The model in $d=1$. All grains from unstable sites in row
$t$ are thrown randomly onto neighboring sites in the next row
$t+1$.}
\label{fig:modelrules}
\end{figure}

Consider a two dimensional square lattice as shown in Fig.
\ref{fig:modelrules}. The direction of propagation is labelled by
$t$, with $0\leq t < T$. The transverse direction is labelled by
$x$, with periodic boundary conditions. On each site, an integer
variable $z(x,t)$ is assigned. The $i$'th grain is added to a
randomly chosen site $x_i$ on the top row $t=0$. There
$z(x_i,0)\rightarrow z(x_i,0) +1$.  When any site acquires a
height greater than $z_c$ it topples, transferring all the grains
at that site, i.e.\ $z(x,t) \rightarrow 0$ for $z(x,t)>z_c$. Each
grain from a toppling site is given equal probability to go to
either downstream nearest neighbor, independent of where the other
grains from the toppling site are placed.  For each toppling
event, the total number of grains are conserved. This is true
except at the open boundary $t=T$ where toppling sites simply
discharge their grains out of the system.

Sites are relaxed according to a parallel update until there are
no more unstable sites, and the properties of the resulting
avalanche are recorded. Then a new avalanche is initiated by
adding a single grain to a randomly chosen site on the top row,
$t=0$.  An avalanche can be characterized by its longitudinal
extent, $t_c$, the largest $t$ row affected, its width, $x_c$, the
largest transverse distance from the avalanche origin to any site
affected by the avalanche, its area, $a$, the total number of
sites affected, its size, $s$, the total number of grains thrown
in toppling events, and the maximum number of grains thrown at a
single site which topples, $n_c$. The fact that $z$ is set to zero
at a toppling site makes the model non-Abelian. The corresponding
rule for the A-SDM for an unstable site is e.g. $z(x,t)
\rightarrow z(x,t) -2$.

In a recent work, Dhar \cite{ADPM} has shown that the stochastic
Manna model \cite{manna}, where a fixed number of grains are
removed from toppling sites, exhibits the Abelian property and is
a special case of the Abelian Distributed Processors Model.  This
property was used to solve analytically for the critical state
properties of the A-SDM, since in that case the Abelian property
makes the appropriately mapped dynamics invertible \cite{pb}.  It
is only necessary to realize that for stochastic models, instead
of associating probabilities with each toppling to determine where
the grains will be thrown, we can assign to each site an infinite
stack of independent, identically distributed random numbers. The
quenched random numbers in each site's stack then determine the
allocation of grains during each toppling event.  Thus, for a
given quenched array, and initial state of the system, the order
of updates will not change the final configuration reached when
the number of grains thrown from an unstable site is fixed.  This
Abelian property makes the directed model invertible, which leads
to the product measure property of the metastable states, and the
solvability of the A-SDM.  However, when all grains from unstable
sites are removed in toppling then the model is not Abelian
anymore.

It is straightforward to generalize the definition of our
non-Abelian sandpile model to higher dimensions, with the number
of directions transverse to the direction of propagation being
$d$.  The threshold $z_c$ can be chosen either to scale with
dimension as $z_c=2d-1$ (for $d \geq 1$) or it can remain constant
at $z_c=1$.  The same behaviour and scaling exponents are
recovered under both conditions.  Below, unless stated otherwise,
we refer to the model with $z_c=1$.

First, we discuss the case $d=1$.  From the dynamical rules it is
clear that the avalanches must, themselves, be essentially
compact. Thus each avalanche sweeps out areas of the lattice
leaving empty sites. At the edges of the avalanche, sites occupied
with grains may remain.  Thus in the stationary state the
structure of the sandpile will consist roughly of empty areas
bounded by wandering paths of occupied sites, which can branch and
recombine.  At the top of the sandpile, where the grains are
added, the network of grains is dense, but pushing into the
sandpile it becomes coarser and coarser.  This coarsening reflects
the fact that avalanches that reach further into the system are
bigger and wider and thus leave traces at their edges that are
further apart.  A steady-state sandpile configuration is shown in
Fig. \ref{fig:1+1dlattice}.

In fact the average density of sites occupied with grains scales
with distance from the top of the pile as $\rho(t) \sim
t^{-\alpha}$, with $\alpha = 0.45 \pm 0.02$. In spite of the
vanishing density in the thermodynamic limit, this network of
grains is essential for maintaining the steady state of SOC,
providing a drainage outlet for grains to be transported from the
top to the bottom of the system. The situation for the A-SDM is
completely different.  In that case, the density of occupied sites
is $1/2$ for $d=1$, and the occupation numbers for sites are
completely uncorrelated, being described by a product measure in
the steady state \cite{pb}.

\begin{figure}
\narrowtext
\includegraphics[width=230pt]{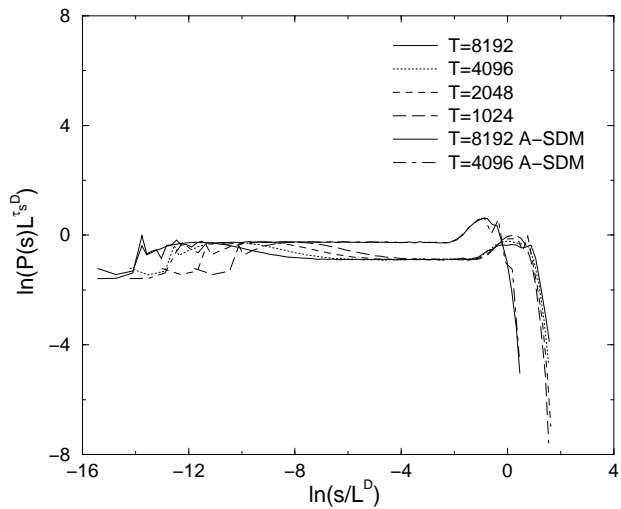}
\caption{Finite size scaling of A-SDM and our sandpile model using
$\tau_s=10/7$, $D=7/4$.} \label{fig:1+1dcollapse}
\end{figure}

Despite these vast differences, the distribution of avalanche
sizes and durations in the steady state of SOC exhibits finite
size scaling with the same critical exponents.  This is
demonstrated in Fig. \ref{fig:1+1dcollapse} where the size
distributions of avalanches, for both the A-SDM and our
non-Abelian model, are presented.  All the critical exponents
characterizing avalanches in the A-SDM have been determined
analytically \cite{pb,maslov} and numerically
\cite{pv,maslov,vazquez}. Avalanches correspond to an absorbing
state phase transition \cite{pmb}.  They are described by a
variant of the Edwards-Wilkinson \cite{ew} stochastic interface
equation where the noise amplitude is a threshold function of the
height (occupancy in the sandpile model) \cite{pb}.  Using these
analytically determined values for critical exponents, good data
collapse is also obtained for our non-Abelian model, where no
analytic solution exists at present.  Thus, within our numerical
accuracy, the critical exponents for the avalanches are the same
in the two cases.  It appears as though the non-Abelain sandpile
has organized large scale structures in such a way as to maintain
the universality class, governed by a stochastic continuum
equation \cite{pb}, for the avalanches.

A configuration in the steady state of our sandpile model in two
dimensions is shown in Fig. \ref{fig:2+1dlattice}.  We observe a
type of domain tube structure with walls separating the different
domains. The tube domains get larger as they go into the system.
Again, a numerical analysis of the avalanche size distribution
using finite size scaling gives critical exponents $\tau_s=3/2$
and $D=2$, the same values as determined analytically for the
A-SDM \cite{pb,maslov}. Similarly for the distribution of
avalanche times, $t$, we find finite size scaling with exponents
$\tau_t=D=2$.  There is no such tube structure, though, in the
A-SDM.

The zero dimensional model is a chain of sites.  Since each site
has only one downstream nearest neighbor, the dynamical rules of
the model must be specified in a slightly different way.  We allow
grains to be distributed to both the nearest and next-nearest
neighbours down the chain, and consider two ways in which the
relaxation of critical sites can be ordered.  With a parallel
update rule all critical sites are relaxed simultaneously. In this
case, the time in terms of the parallel update at which a site can
become critical is not equal to its row number and sites may
topple many times during an avalanche. (This does not occur for $d
> 0$.) We can also define a single-site update rule in which the
top-most unstable site is relaxed each time step. Multiple
toppings cannot occur in that case.  These two cases lead to
different sets of critical exponents for the avalanches in $d=0$.

\begin{figure}
\includegraphics[width=230pt]{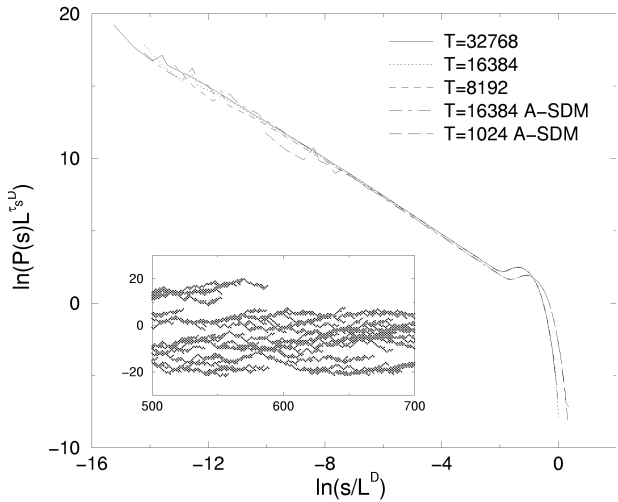}
\caption{Finite size scaling of the distribution of avalanche
sizes for the A-SDM and our model with parallel update in $d=0$,
using $\tau_s=4/3$ and $D=3/2$. Inset shows the position of active
sites away from the average at each time step. }
\label{fig:0+1dcollapse}
\end{figure}

The parallel update dynamics yields the same avalanche exponents,
e.g. $D=3/2, \tau_s=4/3$, as the Abelian model that was studied by
Kloster {\it et al} \cite{maslov} (see Fig.
\ref{fig:0+1dcollapse}).  We found good data collapse for system
sizes ranging from $T\simeq 10^3$ to $T \simeq 3 \times 10^4$, for
both the size, $s$, and time extent, $t$, of the avalanches.
However the average occupancy of sites in the steady state does
not decay to zero as the distance from the top site, where grains
are added, increases, as in our model in higher dimensions.
Instead it is constant $\rho(t)=\frac{1}{4}$, apart from  small
$t$ where the average occupancy adjusts exponentially from the
initial value of $\frac{1}{2}$. Occupied sites do not appear to be
spatially correlated.  Thus the behavior of the model is very
similar to its Abelian relative \cite{maslov} and appears to be in
the same universality class.

In the parallel update dynamics,  as more than one site can topple
at each time step and the location of the topplings is allowed to
vary, the avalanches themselves have structure.  On average the
active front moves through the system with velocity $1.5$ (i.e.\
advances 3 lattice spaces in 2 parallel updates, on average).
Around this average the active sites are split into a series of
smaller fronts which spread, branch and recombine (see inset in
Fig. \ref{fig:0+1dcollapse})

Since our model is not Abelian, a change in the rule for the order
in which unstable sites are updated can have a pronounced effect.
We tried a variety of update rules in higher dimensions $d>0$, but
all the versions we tried appeared to give the same universality
class for the avalanches, although the structure of metastable
states did change drastically.

However, in $d=0$ the critical behaviour is less robust.  A finite
size scaling analysis of the avalanche size distribution using the
single site update rule described above does not collapse with the
same exponents as the $d=0$ A-SDM. A reasonable data collapse can
be produced by changing the cutoff dimension to $D=1.1$ and
keeping $\tau_s=4/3$, but it is not completely convincing. A
multifractal data collapse \cite{kadanoff} did not yield
noticeably better results. The multifractal collapse was performed
with parameters $s_0=0.5$, $l_0=0.2$. For the single site model,
the density of occupied sites does decay going into the system
from the top where the grains are added.  It behaves approximately
as $\rho (s)\simeq 1/t$.

This work was supported by the EPSRC.  We thank P. Bak and S. Lise
for comments on the manuscript.

\end{multicols}

\begin{references}
\bibitem{btw} P. Bak, C. Tang, and K. Wiesenfeld, Phys. Rev. Lett. {\bf 59},
381 (1987).
\bibitem{soc}
For a review see P. Bak,
 {\it How Nature Works: The Science of
Self-Organized Criticality} (Copernicus, New York, 1996).
\bibitem{universal}
 M. Paczuski and S. Boettcher,
 Phys. Rev. Lett. {\bf 77}, 111 (1996).

\bibitem{pv} R. Pastor-Satorras and A. Vespignani, J. Phys. A {\bf 33},
L33 (2000).

\bibitem{pb}M. Paczuski and K. Bassler,
 Phys. Rev. E {\bf 62}, 5347 (2000); the first version of
the paper was published by mistake.  The correct version is web
publication xxx.lanl.gov/abs/cond-mat/0005340. This version
includes a reference to the independent work by Kloster {\it et
al} \cite{maslov}.

\bibitem{maslov}
 M. Kloster, S. Maslov, and C. Tang, Phys. Rev. E, art. no. 026111
 (2001).

\bibitem{ADPM} D. Dhar, Physica A {\bf 270}, 69
(1999); preprint [cond-mat 9902137] (1999).

\bibitem{manna} S. S. Manna, J. Phys. {\bf A 24} L363, (1991).


\bibitem{kadanoff} L. P. Kadanoff, S. R. Nagel, L. Wu, and
S. M. Zhou,  Phys. Rev. A {\bf 39}, 6524 (1989).


\bibitem{tadic}
S. L\"ubeck, B. Tadi\'c, and K. D. Usadel, Phys. Rev. E {\bf 53},
2182 (1996); B. Tadi\'c and R. Ramaswamy, Phys. Rev. E {\bf 54},
3157 (1996); B. Tadi\'c and D. Dhar, Phys. Rev. Lett. {\bf 79},
1519 (1997).

\bibitem{ew} S. F. Edwards and D. R. Wilkinson, Proc. R. Soc. London Ser. A
{\bf 381}, 17 (1982).

\bibitem{vazquez} A. Vazquez, cond-mat preprint 0003420.

\bibitem{pmb} M. Paczuski, S. Maslov, and P. Bak,
Europhys. Lett. {\bf 28}, 295 (1994).

\end{references}
\end{document}